\documentclass{article}
\usepackage{color}
\usepackage{amsmath}
\usepackage{graphicx}
\usepackage{subfigure}

\begin{document}

\title{Quantum States of an Electron Interacting with Various Dielectric Plate Geometries}         
\author{Srihari Sritharan\\University of Pennsylvania\\srisri@mail.med.upenn.edu\\\\Nabil Lawandy\\Department of Engineering\\Brown University}        
\date{\today}          
\maketitle

\begin{abstract}
The tunability of binding energies is explored by modulating a finite dielectric slab width in a planar, three dielectric system. 
After verifying the equivalence of the field method and method of images, three different configurations are explored for possible control of electronic binding energy: A vacuum gap between a Schottky diode, noble gas layers on a metal wall, and an electron confined between two metal plates. In each, varying the width of the finite, middle dielectric was shown to provide control over the binding energy and Bohr radius of the electron. In the case of an electron confined between two plates, it was found that the bound states smoothly connect to the box states as the gap separation was varied. Lastly, forces on the two parallel plates were examined for a possible source of significant repulsion. All numerical calculations were done in MATLAB. 
\end{abstract}

\section*{Introduction}
This paper is focused on the effects of a point charge in a system with three planar dielectrics: a finite dielectric slab between two half-planes. In particular, we strive to determine the tunability of the binding energy, as this provides possibility for practical applications utilizing the corresponding force. There are many ways to actively control the binding energy, such as the amount of charge, or the dielectric constants of the system by controlling temperature. In this study, we focus on controlling the gap width of the finite dielectric slab that interfaces with the two half-planes. We first verify the equivalence of two methods of finding the electron potential from literature: the field method as described in Smythe's book \cite{Smythe} and the method of images \cite{ImagePlane}. We explore three different scenarios to satisfy this goal, the first introducing a vacuum gap between the interface of a Schottky diode to mitigate the effect of band bending. The second configuration places noble gas film layers between a vacuum and metal interface to provide a comparatively smaller binding energy; a possible application of this would be quantum computing as done over helium, but with pulse frequencies for qubit state transitions varying with the number of noble gas layers. The final configuration is an electron confined between two metal plates; we study the bound and box states of the electron and furthermore aim to determine the forces exerted on the plates due to the presence of the electron. A practical application of this study is determining the parameters necessary for the quantum mechanical and electrostatic forces to allow plate levitation and frictionless surfaces. 

\section{Electron Potential in a Three Adjacent  Dielectric Configuration}\label{sec:potentialDerivations}       
In this section, we will derive the field method of finding the potential at the position of an electron, and show equivalence to the method of images. These formulas will be utilized in subsequent chapters for calculations using Schrodinger's equation.
\newcommand{\parenthnewln}{\right.\\&\left.\quad\quad{}}

\subsection{Expansion of Smythe Derivation of Electron in a Finite Dielectric Slab}

Consider three planar dielectrics whose normals are along the z-axis; their respective dielectric constants (relative to vacuum) are $K_1$, $K_2$ and $K_3$. The first dielectric is a half-plane filling the space $-\infty<z<a$, the second is a slab between $a<z<b$, and the third is another half-plane from $b<z<\infty$. 

We place a point charge $q$ at $z_0$ in the second dielectric and observe the potential measured at the location of the point charge due solely to polarization of the dielectrics.

\noindent\makebox[\textwidth]{%
\includegraphics[scale=.41]{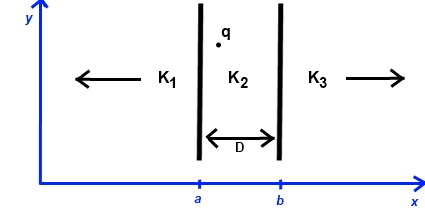}
\label{F:FiniteSlab}
}

First, we consider the Green's function solutions of a point charge in this dielectric configuration. We will work in cylindrical coordinates, such that $z$ is coordinate normal to the interfaces and $\rho$ is the radial component co-planar to the interfaces.\\
The potential in the left half-plane ${-\infty<z<a}$ is solely the contribution from the surface charge on the interface at $z=a$ of $K_1$ and $K_2$.
\begin{equation}
V_1=\frac{q}{4 \pi \epsilon_0 K_2} \int_0^\infty { \phi(k) J_0(k \rho) e^{k(z-z_0)}}\,dk 
\end{equation}
In the region $a<z<b$, the potential contribution is from the polarization at the interface at $z=a$ of $K_1$ and $K_2$, the interface at $z=b$ of $K_2$ and $K_3$, and the Coulomb potential due to the point charge source.

\begin{equation}\label{eq:V2}
\begin{split}
V_2=&\frac{q}{4 \pi \epsilon_0 K_2} \left( \int_0^\infty { \psi(k) J_0(k \rho) e^{-k(z-z_0)}}\,dk + \int_0^\infty { J_0(k \rho) e^{-k|z-z_0|}}\,dk \parenthnewln{+} \int_0^\infty { \theta(k) J_0(k \rho) e^{k(z-z_0)}}\,dk  \right)
\end{split}
\end{equation}
Lastly, the potential in the right half-plane ${b<z<\infty}$ is due to the polarization of the interface at $z=b$ of $K_2$ and $K_3$.
\begin{equation}
V_3=\frac{q}{4 \pi \epsilon_0 K_2} \int_0^\infty { \Omega(k) J_0(k \rho) e^{-k(z-z_0)}}\,dk 
\end{equation}

We need to determine the unknown functions $\phi(k)$, $\psi(k)$ ,$\theta(k)$, and $\Omega(k)$ such that the boundary conditions are satisfied for all $0<=\rho<\infty$. We employ the Fourier Bessel integral to show that if we start with 

\begin{equation}
\int_0^\infty {f_1(k) J_0(k \rho)} \,dk  = \int_0^\infty {f_2(k)J_0(k \rho)} \, dk
\end{equation}

We multiply both sides with $\rho J_0(m \rho) d\rho$ and integrate from $0$ to $\infty$ to get the result

\begin{equation}\label{eq:besselResult}
f_1(m)=f_2(m)
\end{equation}

At each interface, the electric potential and field must match up properly. At the $z=a$ boundary, we require that $V_1=V_2$ and $K_1 \frac{\partial V_1}{\partial z}=K_2 \frac{\partial V_2}{\partial z}$. Similarly at the $z=b$ boundary, the conditions $V_2=V_3$ and $K_2 \frac{\partial V_2}{\partial z}=K_3 \frac{\partial V_3}{\partial z}$ must be satisfied.

Using equation \ref{eq:besselResult} and the above conditions, we arrive at the system of equations:

\begin{equation}
\phi(k) e^{k a'} - \psi(k) e^{-k a'} - \theta(k) e^{k a'} = e^{-k a'}
\end{equation}
\begin{equation}
K_1 \phi(k) e^{k a'} + K_2 \psi(k) e^{-k a'} - K_2 \theta(k) e^{k a'} = -K_2 e^{-k a'}
\end{equation}
\begin{equation}
\psi(k) e^{-k b'} + \theta(k) e^{k b'} - \Omega(k) e^{-k b'} = -e^{-k b'}
\end{equation}
\begin{equation}
-K_2 \psi(k) e^{-k b'} + K_2 \theta(k) e^{k b'} + K_3 \Omega(k) e^{-k b'} = K_2 e^{-k b'}
\end{equation}

where $a'=a-z_0$ and $b'=b-z_0$.

We set this system up as a matrix and solved for the unknown kernels using Cramer's Rule. Since we are concerned about the potential at the location of the point charge, we write out functions $\psi(k)$ and $\theta(k)$, but the other functions can be found similarly.

\begin{equation}
\psi(k) = \frac{1}{\beta_P} \frac{(K_2 - K_1)(K_2-K_3+(K_2+K_3)e^{2 k b}) e^ {-2 k c} } { 1- \frac {\beta_N}{\beta_P} e^{-2 k c}}
\end{equation}

\begin{equation}
\theta(k) = \frac{1}{\beta_P} \frac{(K_2 - K_3)((K_2-K_1)e^{-2 k c}+(K_2+K_1)e^{-2 k b}) } { 1- \frac {\beta_N}{\beta_P} e^{-2 k c}}
\end{equation}

Where we note that $(K_2-K_1)(K_2-K_3)=\beta_N$ and $(K_2+K_1)(K_2+K_3)=\beta_P$. Note that $c=b-a$, equivalent to the gap width $D$ in the figures.
For compact notation, we will rewrite the integrals composing $V_2$ in Eq \ref{eq:V2} as follows,

\begin{equation}
V_2=\frac{q}{4 \pi \epsilon_0 K_2} (I_\psi + I_{source} + I_\theta)
\end{equation}

Considering just the first Bessel integral, substituting in $\psi(k)$,
\begin{equation}
\begin{split}
I_\psi = &\frac{K_2-K_1}{\beta_P} \left( \int_0^\infty { \frac{(K_2-K_3) J_0(k \rho) e^{-k(z-z_0+2 c)}} { 1- \frac {\beta_N}{\beta_P} e^{-2 k c}} }\,dk \parenthnewln{+} \int_0^\infty { \frac{(K_2+K_3) J_0(k \rho) e^{-k(z-z_0-2 a)}} { 1- \frac {\beta_N}{\beta_P} e^{-2 k c}} }\,dk \right)
\end{split}
\end{equation}

Considering just the first term of $I_\psi$, we can expand the denominator as a series

\begin{equation}
\begin{split}
\frac{K_2-K_1}{\beta_P} \int_0^\infty { \frac{(K_2-K_3) J_0(k \rho) e^{-k(z-z_0+2 c)}} { 1- \frac {\beta_N}{\beta_P} e^{-2 k c}} }\,dk &= \frac{(K_2-K_1)(K_2-K_3)}{\beta_P} \left( 
\int_0^\infty {J_0(k \rho) e^{-k(z-z_0+2 c)}}\,dk \parenthnewln{+}\frac{\beta_N}{\beta_P}\int_0^\infty {J_0(k \rho) e^{-k(z-z_0+2 c+ 2c)}}\,dk \parenthnewln{+} \left(\frac{\beta_N}{\beta_P}\right)^2 \int_0^\infty {J_0(k \rho) e^{-k(z-z_0+2 c +4 c)}}\,dk +...   \right)
\end{split}
\end{equation}

We note that $\int_0^\infty{J_0(k \rho) e^{-k|z-z_0|}}=\sqrt{\rho^2+(z-z_0)^2}$  \cite{Smythe} which simplifies the above sum of integrals to 

\begin{equation}
=\frac{(K_2-K_1)(K_2-K_3)}{\beta_P} \sum_{n=0}^\infty {\frac{(\beta_N/\beta_P)^n}{\sqrt{(z-z_0+2c+2nc)^2+\rho^2}}}
\end{equation}

Similarly, we expand all terms of every integral in $V_2$, and collect terms under a common sum. We are interested in the potential at the location of the point charge, so we set $z=z_0$ and $\rho=0$. Due to this, we will disregard the divergent source term $I_{source}$.

\begin{equation}\label{eq:smytheFinal}
V_2=\frac{q}{4 \pi \epsilon_0 K_2} \sum_{n=0}^\infty {\left(\frac{\beta_N}{\beta_P}\right)^n \left(\frac{(\beta_N/\beta_P)}{(n+1)c} + \frac{\beta_{23}}{2nc+2b'} + \frac{\beta_{21}}{2nc+2a''} \right)}
\end{equation}

Where we define $\beta_{21}= \frac{K_2-K_1}{K_2+K_1}$ and $\beta_{23}= \frac{K_2-K_3}{K_2+K_3}$ and $a''=-a'=z_0-a$, such that $a''$ is the positive distance between the point charge and the left half-plane dielectric.

\subsection{Comparison to Images}    

We show here that the above Smythe approach provides the same analytical answer as that derived from the method of image charges. 
We arrive at two sequences of distance-image charge pairs, depending on the side chosen for the initial image interaction.

\begin{figure}[!htbp]
\caption{First few images if initial image charge is in $K_1$}
\centering
\includegraphics[scale=.6]{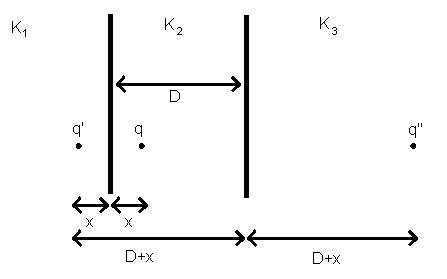} 
\label{F:Images}
\end{figure}

Supposing the first image charge to be in $K_1$, we get the following sequence:

\begin{equation}\label{eq:positions1}
\lbrace(r_i,q_i)\rbrace_{i=0}^{\infty}=\lbrace (2 i c + 2a'', \beta_{21} ^{i+1}\beta_{23} ^{i} q) \rbrace_{i=0}^{\infty} \cup \lbrace (2 i c +2c, \beta_{21} ^{i+2}\beta_{23} ^{i+2} q) \rbrace_{i=0}^{\infty}
\end{equation}

And similarly if the initial image charge is in $K_3$, 
\begin{equation}\label{eq:positions2}
\lbrace(r_j,q_j)\rbrace_{j=0}^{\infty}= \lbrace (2 j c + 2b', \beta_{23} ^{j+1}\beta_{21} ^{j} q) \rbrace_{j=0}^{\infty} \cup \lbrace (2 j c +2c, \beta_{23} ^{j+2}\beta_{21} ^{j+2} q) \rbrace_{j=0}^{\infty}
\end{equation}

Finally, we can compute the sum of all image charge Coulomb potential contributions at $z_0$, and by comparing terms we note that indeed Eq \ref{eq:smytheFinal} and the below are equal, observing that $\beta_{21}\beta_{23}=\frac{\beta_N}{\beta_P}$.

\begin{equation}\label{eq:slabImages}
V_2=\frac{q}{4 \pi \epsilon_0 K_2} \left(\sum_{i=0}^{\infty} \frac{q_i}{r_i} + \sum_{j=0}^{\infty} \frac{q_j}{r_j}  \right)
\end{equation}

\subsection{Limiting Cases}
We wish to ensure that our equations produce reasonable behavior in analytically known limiting cases. 
First, suppose two adjacent dielectrics have equal relative permittivity; as an example we set $K_1=K_2$. This results in $\beta_{N}=0$ and $\beta_{P}=2 K_2 (K_2 + K_3)$. This significantly simplifies Eq \ref{eq:smytheFinal} to give us:

\begin{equation}
V_2=\frac{q}{4 \pi \epsilon_0 K_2} \frac{K_2-K_3}{K_2+K_3} \frac{1}{2 b'} = \frac{q}{4 \pi \epsilon_0 K_2} \frac{\beta_{23}}{2 b'}
\end{equation}

which we recognize as the image charge method result of an electron in $K_2$ facing an half-plane of $K_3$.
We can observe a further simplification to check an electron in vacuum facing a metal: $K_2=0$ and $K_3=\infty$, which yields

Another case we seek to test is the limit as $b$, the position of the $K_3$ half-plane dielectric, approaches infinity. We expect to see an match with the image method result of an electron in $K_2$ facing an half-plane of $K_1$ on its left. We begin examining this limiting case by breaking apart this series in Eq \ref{eq:smytheFinal}, taking out the first ($n=0$) term of each sum. This expands to 
\begin{equation}
V_2=\frac{q}{4 \pi \epsilon_0 K_2} \left[ \left(\frac{(\beta_N/\beta_P)}{c} + \frac{\beta_{23}}{2b'} + \frac{\beta_{21}}{2a''} \right) + \sum_{n=1}^\infty {\left(\frac{\beta_N}{\beta_P}\right)^n \left(\frac{(\beta_N/\beta_P)}{(n+1)c} + \frac{\beta_{23}}{2nc+2b'} + \frac{\beta_{21}}{2nc+2a''} \right)} \right]
\end{equation}

Since $b$ is infinity, so is $c$, which simplifies the above to:
\begin{equation}
V_2=\frac{q}{4 \pi \epsilon_0 K_2} \frac{\beta_{21}}{2a''}
\end{equation}

which is exactly the result we expected.

\subsection{Expansion of Smythe Derivation of Electron in Dielectric Half-Plane}\label{S:leftHalfPlane}
We will again consider the same three dielectric setup described at the beginning of the previous section, but instead we now place the potential at a point charge placed in the half-plane of $K_1$. We use the same Smythe approach as above, but with different initial integrals for the potential at each region, moving the source term to $V_1$. This gives us the following altered system of equations:

\begin{equation}
\phi(k) e^{k a'} - \psi(k) e^{-k a'} - \theta(k) e^{k a'} = -e^{-k a'}
\end{equation}
\begin{equation}
K_1 \phi(k) e^{k a'} + K_2 \psi(k) e^{-k a'} - K_2 \theta(k) e^{k a'} = K_1 e^{-k a'}
\end{equation}
\begin{equation}
\psi(k) e^{-k b'} + \theta(k) e^{k b'} - \Omega(k) e^{-k b'} = 0
\end{equation}
\begin{equation}
-K_2 \psi(k) e^{-k b'} + K_2 \theta(k) e^{k b'} + K_3 \Omega(k) e^{-k b'} = 0
\end{equation}

For brevity, we will not show the remainder of the derivation, but rather state the final form of the potential at the point charge:

\begin{equation}\label{eq:leftPlaneFinal}
V_1=\frac{q}{4 \pi \epsilon_0 K_1} \frac{1}{\beta_P} \sum_{n=0}^\infty {\left(\frac{\beta_N}{\beta_P}\right)^n \left( \frac{\beta_{D}}{2nc+2a} - \frac{\beta_{C}}{2nc+2b} \right)}
\end{equation}

Where $\beta_C=-K_2^2+K_2 K_3 -K_1 K_2 +K_1 K_3$ and $\beta_D= -K_2^2-K_2 K_3 +K_1 K_2 +K_1 K_3$. Since the point charge is always to the left of the two interfaces, we denote $a$ and $b$ as the positive distance of the point charge to the first and second interfaces, respectively.

\subsection{Limiting Cases}
We will again address two limiting cases to examine the derived series for reasonable behavior. Consider if $K_1=K_2$, effectively removing the $K_2$ dielectric, we expect to see the image method result of an electron in $K_1$ facing a wall of $K_3$ at distance $b$. 

We break up the series like before, extracting the first $(n=0)$ term from the series:
\begin{equation}
V_1=\frac{q}{4 \pi \epsilon_0 K_1} \frac{1}{\beta_P} \left[ \left( \frac{\beta_{D}}{2a} - \frac{\beta_{C}}{2b} \right) + \sum_{n=1}^\infty {\left(\frac{\beta_N}{\beta_P}\right)^n \left( \frac{\beta_{D}}{2nc+2a} - \frac{\beta_{C}}{2nc+2b} \right)} \right]
\end{equation}

Setting $K_1=K_2$, we note that$\beta_{N}=0$, $\beta_{D}=0$, leaving behind just:
\begin{equation}
V_1=\frac{q}{4 \pi \epsilon_0 K_1} \frac{\beta_{C}}{\beta_P} \frac {1}{2a}
\end{equation}
\begin{equation}
V_1=\frac{q}{4 \pi \epsilon_0 K_1} \frac{K_1-K_3}{K_1+K_3} \frac{1}{2a}
\end{equation}

And we recover the image solution as expected.

On the other hand if we set $c=\infty$, we expect to see the image method result of an electron in $K_1$ facing a wall of $K_2$ at distance $a$. As before, we extract the first term of the series:
\begin{equation}
V_1=\frac{q}{4 \pi \epsilon_0 K_1} \frac{1}{\beta_P} \left[ \left( \frac{\beta_{D}}{2a} - \frac{\beta_{C}}{2b} \right) + \sum_{n=1}^\infty {\left(\frac{\beta_N}{\beta_P}\right)^n \left( \frac{\beta_{D}}{2nc+2a} - \frac{\beta_{C}}{2nc+2b} \right)} \right]
\end{equation}
We set $c=\infty$ and $b=\infty$ , and we see many terms of the potential go to zero, leaving just:
\begin{equation}
V_1=\frac{q}{4 \pi \epsilon_0 K_1} \frac{\beta_{D}}{\beta_P} \frac{1}{2a}
\end{equation}

If we finally say that the $K_3$ is effectively so far away its contribution does not exist, we set $K_3=0$ and we see that we recover the image solution:
\begin{equation}
V_1=\frac{q}{4 \pi \epsilon_0 K_1} \frac{K_1-K_2}{K_1+K_2} \frac{1}{2a}
\end{equation}

\section{Dielectric Interfaces}
Our main goal is to use the Schrodinger equation to study electronic states for a given dielectric configuration and understand the behavior of the Bohr Radii and binding energies as we vary the dielectric constants. We strive to find configurations that provide strong binding energies (and forces). We know from literature that a single electron in a vacuum facing a metal wall experiences a binding energy of -0.85eV \cite{ImagePlane}. Based on this information, we estimate that we could get desirable results with just two dielectrics for our configuration.

\noindent\makebox[\textwidth]{%
\includegraphics[scale=.55]{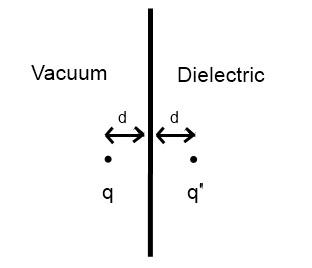}
\label{F:MetalWall}
}

We consider simple image state example - an electron in vacuum facing a half plane dielectric. Suppose the charge is a distance $d$ from the plane and has charge $q$.
From the method of image charges, we know that we can place an image charge in the dielectric at a distance $2d$ away from the point charge, and that is representative of the surface charge contribution that the point charge sees. The magnitude of the point charge is dependent on the dielectric constants of half-plane:

\begin{equation}
q' = \frac{1-\epsilon}{1+\epsilon} q 
\end{equation}

With this configuration, we can consider the point charge to be facing a Hydrogen-like atom \cite{Lifshitz}, at a distance $2d$ away from the nucleus (here, our image charge).

\begin{eqnarray}\label{eq:simpleImage}
V= Z\frac{q q'}{r} = \frac{1-\epsilon}{1+\epsilon} \frac{q^2}{4d} \\
Z= \frac{1}{4} \frac{1-\epsilon}{1+\epsilon}
\end{eqnarray}

We solve Schrodinger's equation analytically with this potential over the domain $(-\infty,0]$ to get the wavefunctions.

We note here that although the potential from classical electrostatics is given by $V \propto 1/(2d)$, we use the potential $V \propto 1/(2*2d)$ with the extra factor of 1/2 when we apply it in Schrodinger's equation. This is because the potential is instead derived from the work necessary to bring the charge from infinity in the present field. 
In the case of a simple Coulomb attraction:
\begin{equation}
\int_{-\infty}^z \frac{-q^2}{4 \pi \epsilon} \frac{1}{z'^2} dz' = -\frac{1}{2} \frac{q^2}{4 \pi \epsilon} \frac{1}{z}
\end{equation}

Therefore similarly, we solve for the eigenvalues in our configuration by incorporating a factor of $1/2$ into the valency of Eq \ref{eq:simpleImage}.

The first few radial wavefunctions are nearly identical to their hydrogen counterparts:

\begin{eqnarray}
&R_1 = &2 \left(\frac{Z}{a_0}\right) ^ {\frac{3}{2}} e ^ {-Zr/a_0} 
\\&R_2 = &2 \left(\frac{Z}{2a_0}\right) ^ {\frac{3}{2}} \left(1-\frac{Z}{2a_0}\right) e ^ {-Zr/{2a_0}}
\\&R_3 = &2 \left(\frac{Z}{3a_0}\right) ^ {\frac{3}{2}} \left(1-\frac{Z}{3a_0}+ \frac{2(Zr)^2}{27a_0^2}\right) e ^ {-Zr/{3a_0}}
\end{eqnarray}

whereas here, we use the valency magnitude $Z=\frac{1}{4}(\epsilon-1)/(\epsilon+1)$, in contrast to $Z=1$ for hydrogen.

The respective energy levels for the above wavefunctions are given by:

\begin{equation}
E_n= - \frac{Z^2 m e^4}{2 \hbar^2} \frac{1}{n^2} =  - \left(\frac{1-\epsilon}{1+\epsilon}\right)^2 \frac{ m e^4}{32 \hbar^2 n^2}
\end{equation}

The corresponding Bohr radius for these energies are given by:
\begin{equation}
Bohr_n = n^2 \frac{a_0}{Z} = 4 n^2 \frac{\hbar^2}{m e^2} \frac{\epsilon+1}{\epsilon-1}
\end{equation}

where $a_0$ is the Bohr radius for hydrogen, given by $a_0= \hbar/(m e^2)$.

Therefore, the electron in vacuum facing a wall of metal would experience a ground state energy of -0.85eV and a Bohr radius of 0.2116 nanometers, 1/16 times and 4 times their Hydrogen counterparts, respectively. 

In general, we can employ a two dielectric interface with a vacuum and a semiconductor. Quantum computing with electrons on liquid helium is a current application of this two dielectric interface \cite{Helium}. Here, an electron is confined to a potential well above liquid helium, created by the attraction to the image charge and the repulsion from the Helium electrons. The electron resides at some considerable height above the Helium, at 11nm for the ground state and 46nm for the first excited state. By applying a microwave pulse at a frequency $f_R = (E_2 - E_1)/h = 120$ GHz, the electron can be boosted from the ground state to the first excited state. \cite{Helium} By operating between these two "0" and "1" states, this system operates as a qubit; that is, a binary computation unit at the quantum scale.

We quickly find however that having just two dielectrics in contact is not as desirable as we had hoped. Consider an example Schottky diode \cite{Schottky}, a left half-plane n-doped semiconductor interfacing directly with a right half-plane metal. The problem with working in this configuration, however, is that due to the excess of electrons in the metal and the excess of holes in the semiconductor, electrons flow from the metal to the semiconductor, causing a built-in positive potential across the depletion region, which diminishes the depth of the potential well formed by the electron facing the metal. As a result of this weaker potential well, we recover considerably weaker binding energies of the electron as well. With an electron in vacuum facing a wall of metal, we recover the ground state binding energy of -0.85 eV, but with the electron in gallium arsenide (GaAs) facing a wall of metal, we only recover -.00049 eV.

This is the 'band bending' phenomenon \cite{Schottky}, where the electrons transfer across the interface to equilibrate the Fermi levels of each material. The motion of electrons is at equilibrium when there is a balance between the diffusion force (modulated by the diffusion constant $D_0$) and the electric field force:
\begin{equation}
D_0 \Delta \rho = \sigma E
\end{equation}

\subsection{Finite Vacuum Gap}
We can ameliorate serious loss of binding energy from band bending by not letting the dielectrics come in direct contact. Consider a dielectric left half plane, followed by a finite vacuum gap, and then a metal right half plane. This way, the rate of tunneling across the vacuum is greatly reduced, though not completely nullified. 

\noindent\makebox[\textwidth]{%
\includegraphics[scale=.55]{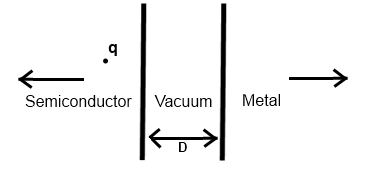}
\label{F:VacuumGap}
}

We examined the effect of modulating the vacuum gap size with a left half-plane semiconductor, using GaAs and InSb as our examples. GaAs has a comparatively larger dielectric constant of 12.9 \cite{GaAs} and InSb has a dielectric constant of 16.8 \cite{InSb}. We were interesting in examining the effects of modulating the vacuum gap width.

Since the electron is present in a semiconductor, its effective mass and corresponding hole mass are smaller than that of an electron in vacuum; effective electron mass and hole mass is $0.67e$ and $0.45e$ for GaAs and $0.013e$ and $0.6e$ for InSb, respectively\cite{GaAsEffectiveMass}\cite{InSbEffectiveMass}. We note that we use the potential function derived in Section \ref{S:leftHalfPlane}.
We proceed to calculate the ground state energy of the electron for varying vacuum gap widths, as well as plot the wavefunction for increasing vacuum widths to observe the effect of moving the metal plate further away.

\noindent\makebox[\textwidth]{%

\includegraphics[scale=.41]{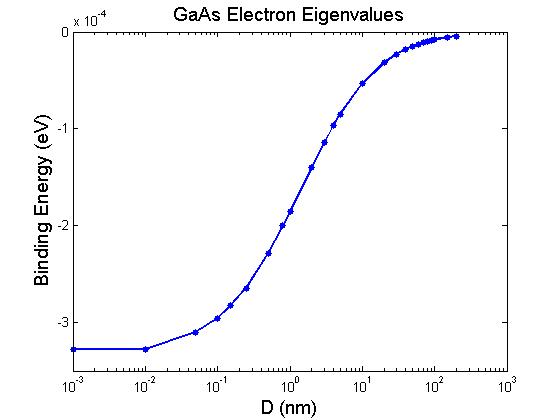}
\label{F:GaAsElectron}

\includegraphics[scale=.41]{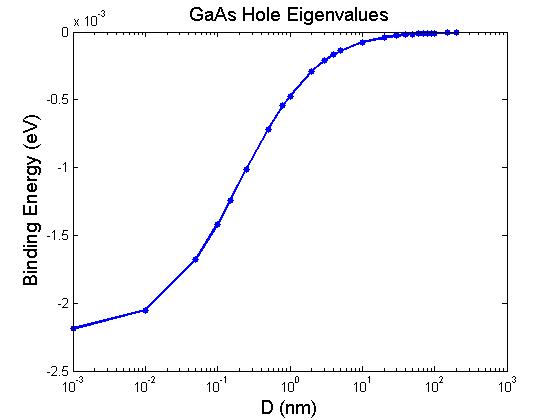}
\label{F:GaAsHole}
}

\noindent\makebox[\textwidth]{%

\includegraphics[scale=.38]{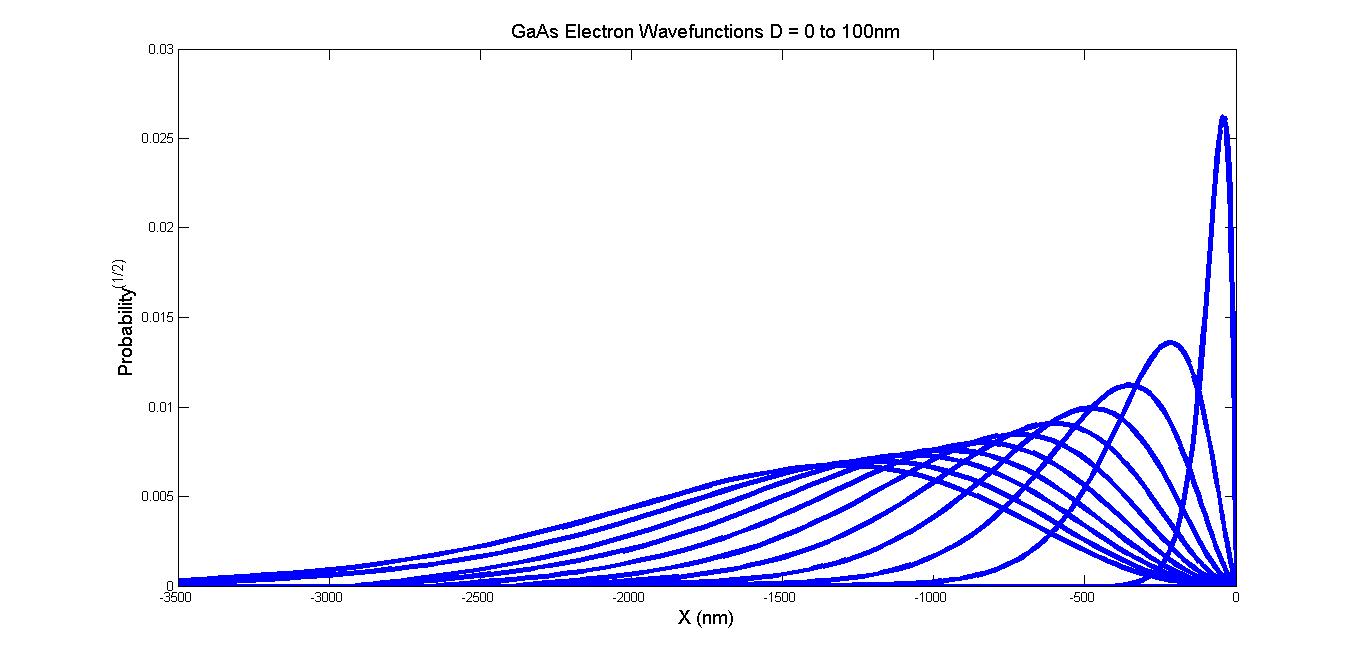}
\label{F:GaAsElectronPsi}
}
\noindent\makebox[\textwidth]{%

\includegraphics[scale=.38]{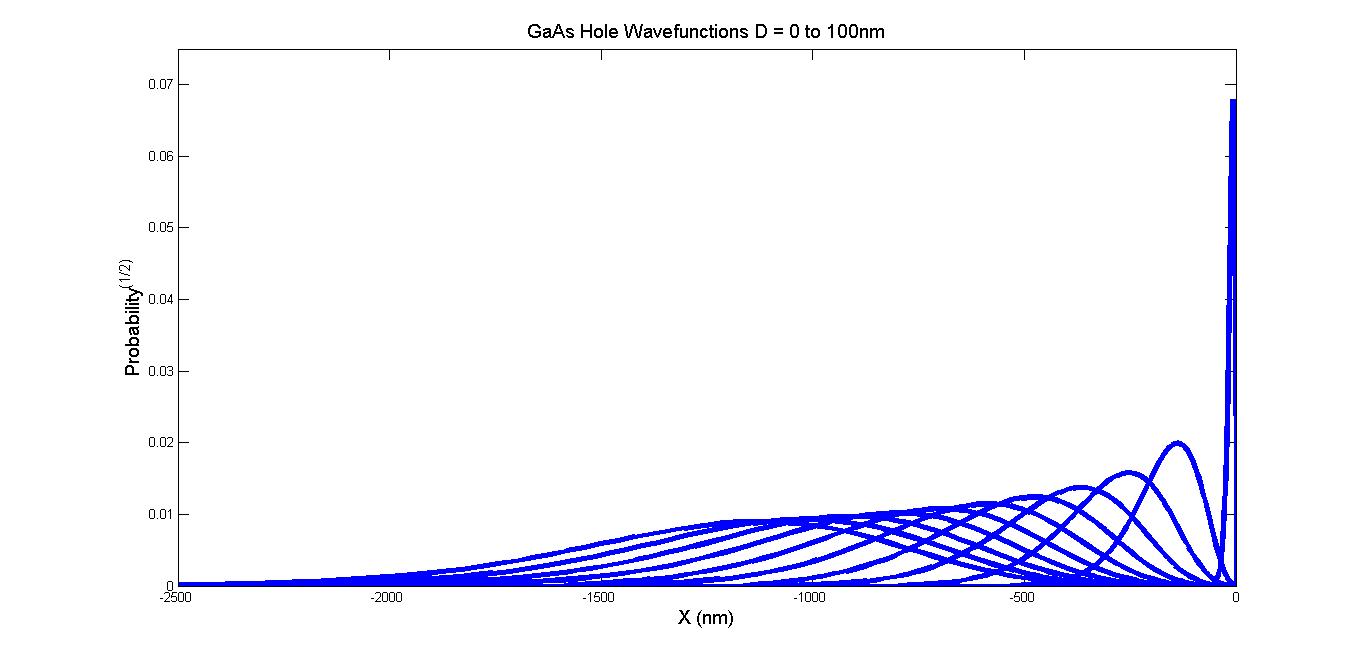}
\label{F:GaAsHolePsi}
}

\noindent\makebox[\textwidth]{%

\includegraphics[scale=.41]{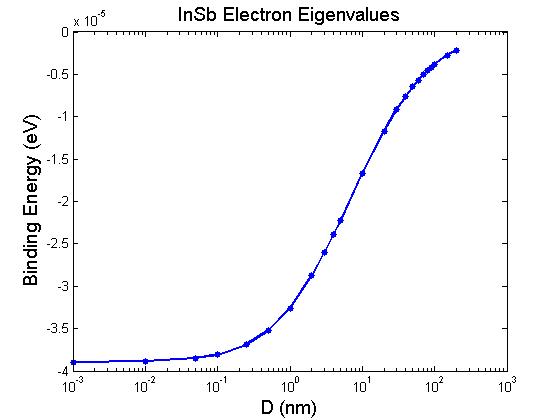}
\label{F:InSbElectron}

\includegraphics[scale=.41]{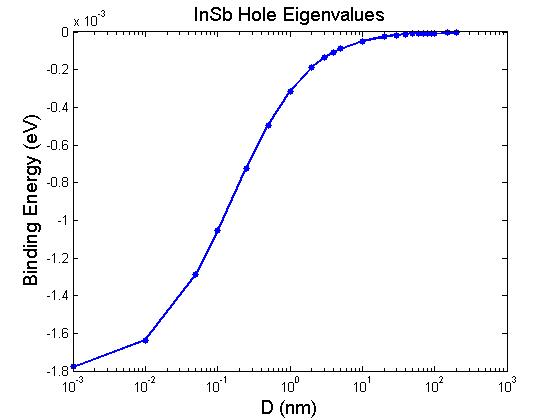}
\label{F:InSbHole}
}

\noindent\makebox[\textwidth]{%

\includegraphics[scale=.38]{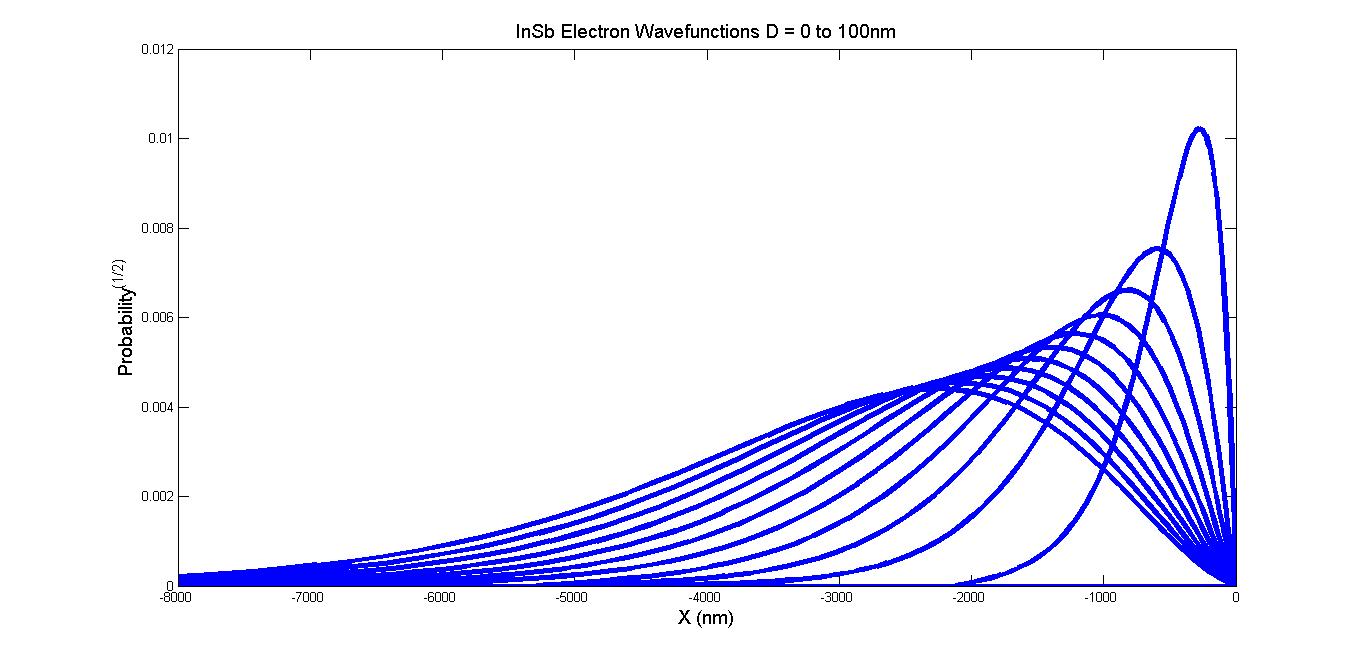}
\label{F:InSbElectronPsi}
}

\noindent\makebox[\textwidth]{%

\includegraphics[scale=.38]{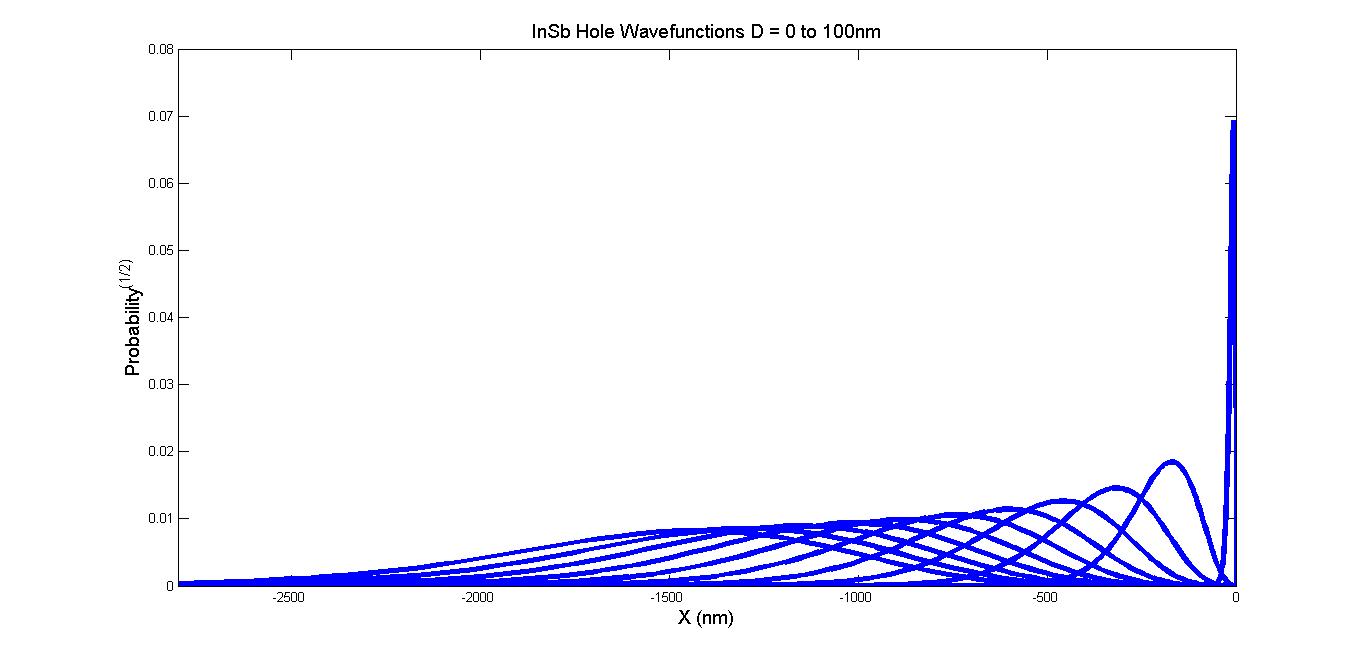}
\label{F:InSbHolePsi}
}

As we expected, we are able to finely tune the Bohr radius and binding energy of the electron, from the limiting case of a Schottky diode of GaAs or InSb, to an infinitely distant metal plate. This control is particularly due to our ability to increase the vacuum gap width over a continuous range from zero to infinity.

In the graphs of the wavefunctions, the vacuum gap was increased from 0 to 100nm in increments of 10nm. We can see the broadening of the probability mass and a consequent shift of the Bohr radius away from the interface. To keep the scale of the graph at a reasonable viewing ratio, we plotted the square root of the probability; however, the proportionality of each individual wavefunction is intact and is still representative of the probability of locating the electron. It should be noted that since the electron mass is in the left half-plane, the interface with the vacuum gap is at $Z=0$ on the right edge of the wavefunction graphs.

To double check that our energy calculations reach the right limiting cases, we can recover the vacuum-metal ground state of $-0.85eV$ that we have discussed previously. The binding energy is proportional to the effective mass and inversely proportional to the square of the dielectric constant. Thus the following equality should hold true: $E_{0 semi} / m_{eff} * (\epsilon_{semi})^2= -0.85eV$, and we find that indeed it does for our calculations.

By looking at figures \ref{F:GaAsElectron},\ref{F:GaAsHole},\ref{F:InSbElectron}, and \ref{F:InSbHole}, we see the energy level drop to near zero as the metal is moved further away from the electron. It is interesting to note that the GaAs - Vacuum interface is in fact, repulsive (GaAs has a larger dielectric constant than the vacuum) and the electron would not be bound. However the presence of the metal, even if very far away, continues to provide an attractive force. This is under the assumption that the area of the plate is so large compared to our gap distances that the electron will not see the edges of the metal plate. If this were not so, the attractive force of the induced, opposite-sign surface charge would be mitigated by the electrostatic repulsion from the same-sign surface charge induced at the edges and corners of the plate to maintain zero total charge on the plate.

\newpage
\subsection{Finite Gap Filled With Noble Gas}
Consider instead that the electron is in a half plane vacuum, followed by a finite semiconductor, and then a wall of metal. Here, we can talk about tunability of the binding energies, Bohr radii, and effective dielectric constant. A classic example that we will use as our finite gap is liquid helium. 

\noindent\makebox[\textwidth]{%
\includegraphics[scale=.55]{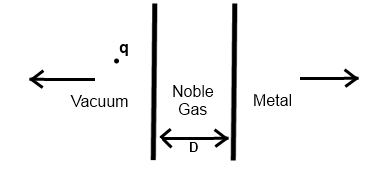}
\label{F:NobleGas}
}

We expect that varying the thickness of the helium will modulate the binding energy and Bohr radius of the electron. Indeed from the following graphs, we can see that the energy and Bohr radius can be varied to the two extremes of either an effective wall of helium or an effective wall of metal. It should be noted however, that unlike the previous example with a vacuum gap, liquid helium cannot be added as any fractional thickness; liquid helium increases thickness in discrete layers of thickness approximately 20nm \cite{HeliumThickness}. Thus, our graphs increase in integer multiples of this discrete thickness.

\noindent\makebox[\textwidth]{%

\includegraphics[scale=.4]{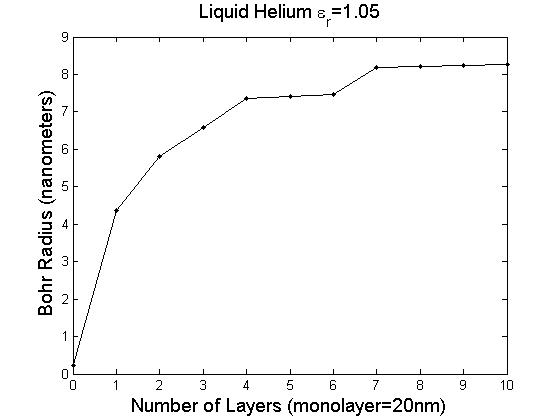}
\label{F:HeliumBohr}

\includegraphics[scale=.4]{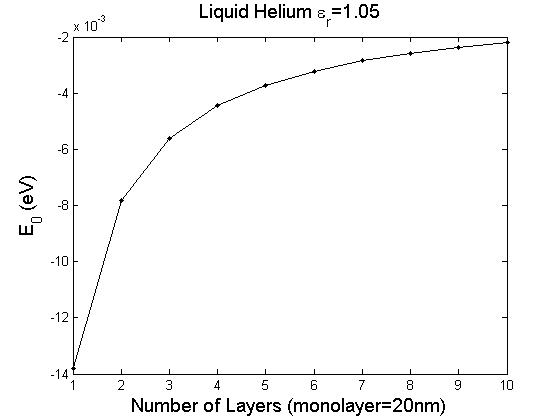}
\label{F:HeliumE0}
}

These graphs show that we can tune the energy and Bohr radius by varying the thickness of the helium slab, and this is an improvement from the previous situation of having either just an electron in a vacuum facing a metal or an electron in a semiconductor facing a metal but with a low binding energy. As a note, the Bohr radius graph does not appear to exhibit smooth data; however we assure that this is due to numerical error. The Bohr radius is calculated as the location of maximum probability, which in turn is affected by the discretization of the domain when the Runge-Kutta solvers were employed. We expect that the real curve will not exhibit any kinks. Examining the $E_0$ vs. $D$ graph, we see that adding just one layer of helium attenuates the binding energy to a little less than $1/50^{th}$ of the original metal binding energy of $-0.85eV$. 

Keeping in mind that a metal has an infinite dielectric constant and that liquid helium's is on the order of $10^0$ (around 1.05) \cite{HeliumEpsilon}, we can attempt to tune the effective dielectric constant that the electron experiences. By varying the thickness of the helium layer from zero to infinity, we change the binding energy of the electron and thus also the effective dielectric constant of the potential experienced by the electron. In the two extreme cases, the electron is facing either a wall of metal or a wall of liquid helium. The effective epsilon was calculated by first finding the curve of Bohr Radius vs. epsilon of an electron in vacuum facing a single wall of $\epsilon$, ranging from $\epsilon_0$ to $\infty$. Then, we found the Bohr radius vs. gap width for the three dielectric case, and used table lookup on the previous curve to determine the effective dielectric constant, as if we were to replace the three dielectrics with a two dielectric analogue. 

\noindent\makebox[\textwidth]{%
\includegraphics[scale=.5]{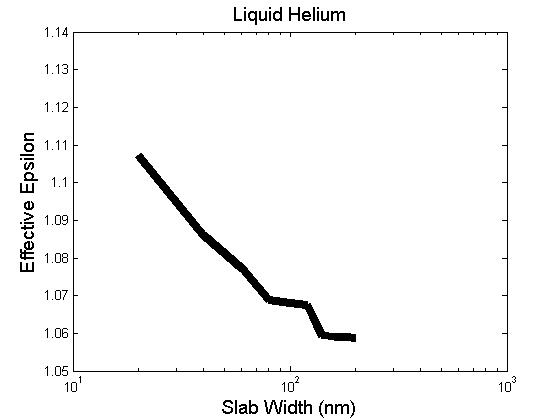}
\label{F:HeliumEffectiveEpsilon}
}

We again note that after the addition of just one layer of helium, the effective dielectric constant drops considerably closer to $1.05$, the dielectric constant of helium. We see that the tunable dielectric constant range is only between $1.1$ and $1.05$. Taking this into consideration, we try another related noble gas film - solid argon. This has a dielectric constant of $1.7$, and a film thickness of $0.345nm$, considerably smaller than liquid helium. Solving for the same properties above, we generate the following graphs. This time, we computed the energy levels and Bohr radii for the ground state (black) and first (blue) and second (red) excited states.

\noindent\makebox[\textwidth]{%

\includegraphics[scale=.4]{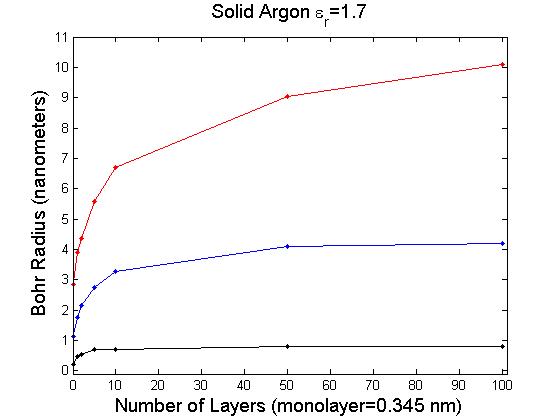}
\label{F:ArgonBohr}

\includegraphics[scale=.4]{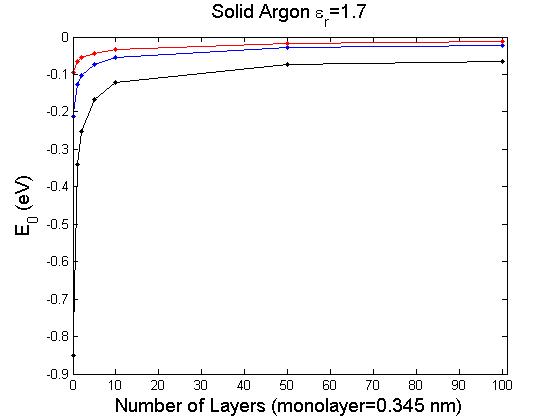}
\label{F:ArgonE012}
}

This time, the results are more promising, as the ground state reduces only a little less than half in ground state energy, providing a more refined control of the system by increasing the number of layers of argon. Once more we calculate the effective dielectric constant that the electron would face if the system was only two dielectrics:

\noindent\makebox[\textwidth]{%
\includegraphics[scale=.5]{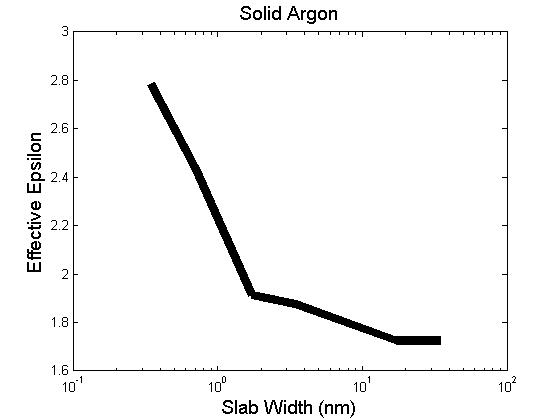}
\label{F:ArgonEffectiveEpsilon}
}

In line with our analysis of the previous graphs, we see that adding one layer does not decrease the effective dielectric constant too close to the constant of argon ($1.7$), and provides a tunable range between $2.78$ and $1.7$.

\section{Considerations of a Quantized Particle Between Two Plates}\label{S:particlePlates}
When a particle is placed in the gap between two plates, it is interesting to not only study the behavior of the particle, but also the forces on the plates themselves. A simple mass between the plates, such as a single neutron, feels a particle-in-a-box-like potential, where $V=0$ in the gap, and $V=\infty$ at the boundary "walls." It should be noted that this is not a consequence of electrostatics, but a quantum phenomenon.

\noindent\makebox[\textwidth]{%
\includegraphics[scale=.3]{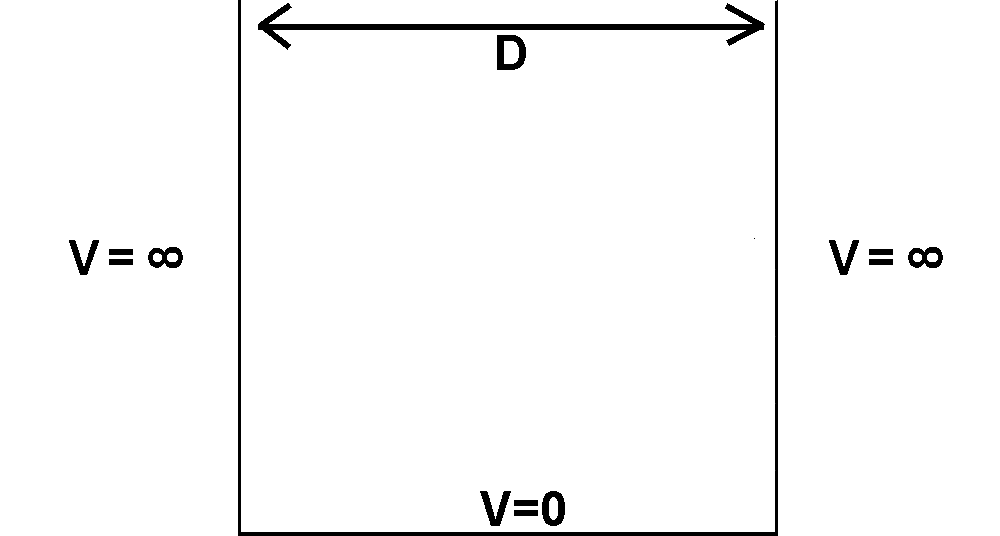}
\label{F:ParticleInABoxPotential}
}

The energy at each quantum number is given by:
\begin{equation}
E_n = \frac{n^2 \hbar ^2 \pi ^2}{2 m L^2}
\end{equation}

where $L$ is the distance between the plates, and $m$ is the mass of the particle. We can calculate the repulsive force felt by the plates by taking the partial derivative with respect to $L$:

\begin{equation}
F_n= -\frac{\partial{E_n}}{\partial{L}}= - \frac{n^2 \hbar^2 \pi^2}{m L^3}
\end{equation} 

Although we did not previously specify the orientation of the plates, we now configure such that one plate is set flat on the Earth and the other plate is set on top, such that the plate normal vectors are collinear with gravity. An intriguing question now arises; can the particle force on the upper plate counteract the force of gravity? Setting the above equation equal to the gravitation force provides:

\begin{equation}
F_n + M g = 0
\end{equation}
\begin{equation}
M= - \frac{n^2 \hbar^2 \pi^2}{m L^3 g}
\end{equation}

where $M$ is the mass of the plate. This equation allows us to determine how much mass $M$ we can levitate at a distance $L$ from the plate at a specific energy level $n$. A specific example can be constructed with a neutron particle mass and $n=1$ for the zero-point energy. With these parameters:

\begin{eqnarray}
m = 1.67 * 10^{-27} kg \\
n = 1
\end{eqnarray}

we arrive at the equation $M =(6.7022 * 10^{-42})(L^{-3}) $. Even for a modest levitation of $L = 1$ nm, we get a mass of only $M = 6.7022 * 10^{-15}$ kg, which is about the mass of a single bacterium.

It suffices to say that a mass alone in a square potential well does not provide enough force to produce a significant levitation of the plate. Thus, we turn to utilizing charged particles that can polarize the plates and provide additional, electrostatic repulsion.

\subsection{Electron Between Two Infinite Dielectric Constant Plates}
An electron between two plates induces surface charge on each plate, resulting in a double well attractive potential at the electron, as well as a repulsive force between the two metal plates. Compared to a single well potential, double well potentials are especially interesting in that the bound states are wavefunctions that come in pairs, presented as a symmetric and an antisymmetric wavefunction whose corresponding eigenvalues are close in value. 

\noindent\makebox[\textwidth]{%
\includegraphics[scale=.35]{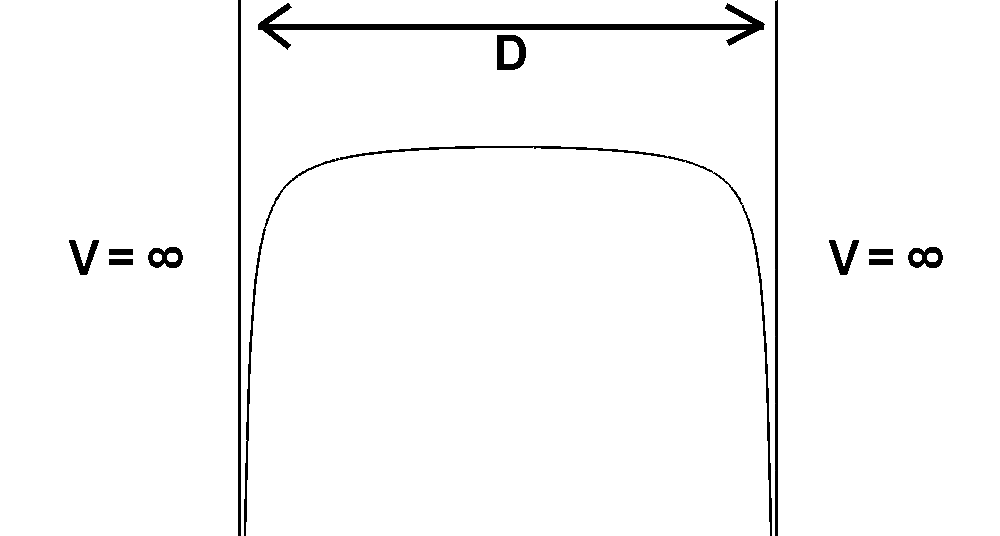}
\label{F:DoubleImageWell}
}

To explore the effect of narrowing the distance between the plates, we plotted the total potential energy for various gap widths. The most blue curve is the narrowest gap and the most red curve is the largest gap. We see that as the the plates come closer together, the potential value at the center becomes increasingly negative, indicating a possibly lower energy necessary for the electron to escape the well and exhibit box states. We discuss this further in section \ref{S:Schro} when we generate the electronic binding energy curve.

\noindent\makebox[\textwidth]{%
\includegraphics[scale=.35]{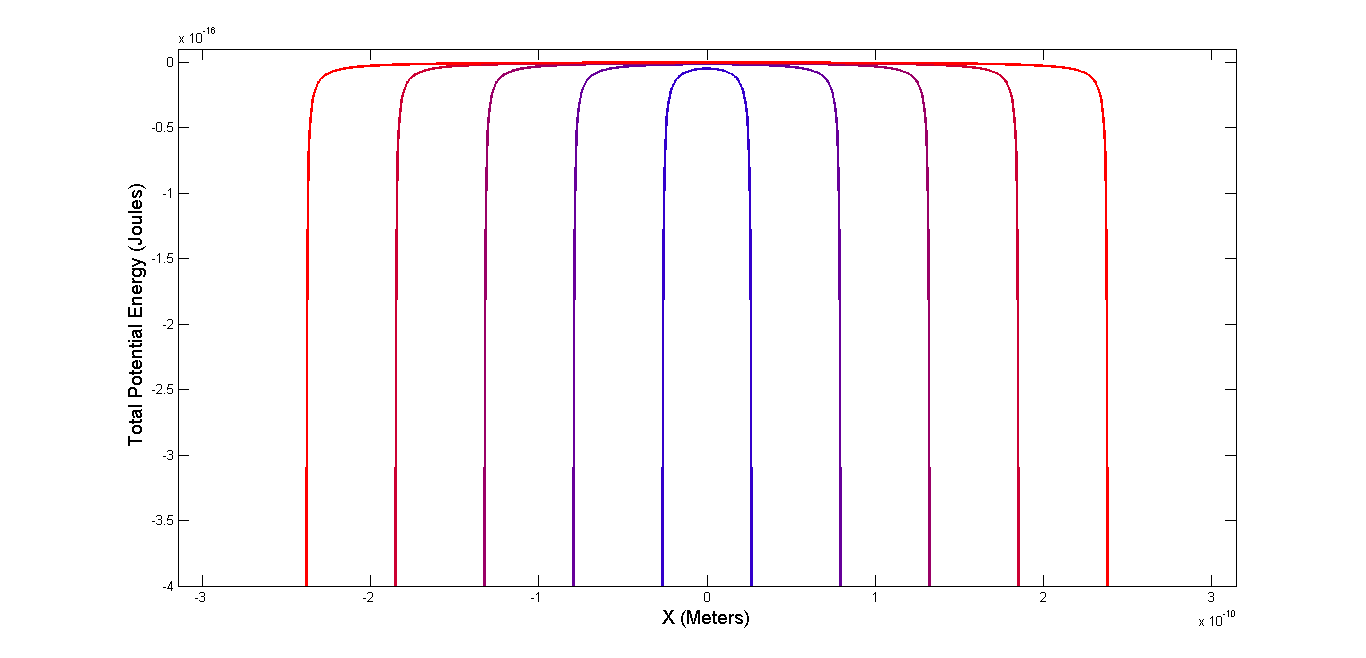}
\label{F:TotalPotentialVariousGaps}
}

Ammonia, chemical compound $NH_3$, is a pyramidal molecule with a double well analogue; the three hydrogens form a flat trianglar plane and the nitrogen N atom can take the apex on either side of the plane to form the pyramid. 
In the below figure, the double well is drawn along with the first symmetric-antisymmetric energy pair. 

\noindent\makebox[\textwidth]{%
\includegraphics[scale=.35]{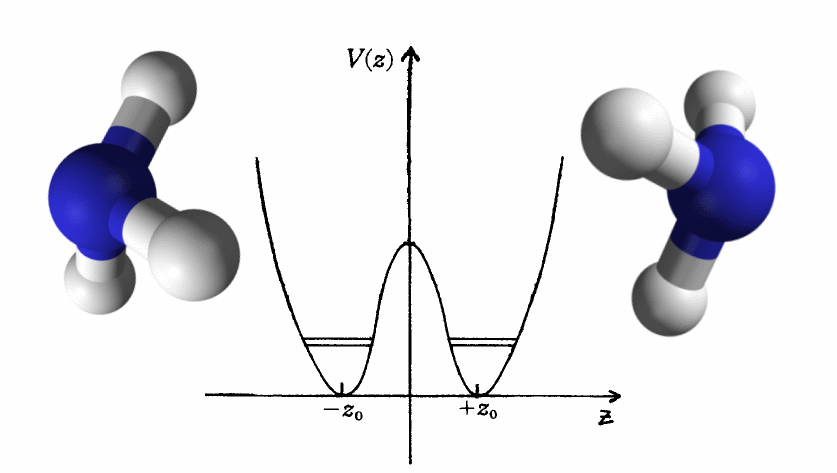}
\label{F:InversionOfAmmonia}
}
In the ball-stick model images, the blue atoms represent nitrogen and the grey ones represent hydrogen. \cite{InversionAmmonia}

Vibrational energy can perturb the N atom from one well until the nitrogen molecule is coplanar with the hydrogens (as seen by the local maxima in the potential graph at $z=0$). At this unstable equilibrium, the nitrogen molecule can then settle into the other well, known as ammonia inversion. \cite{Ammonia} An important application of this configuration inversion is the ammonia maser (Microwave Amplification by Stimulation Emission of Radiation), a pioneering work by Charles Townes and his students at Columbia University \cite{Maser}. This device exploits the symmetric and antisymmetric states of the molecule by applying an inhomogenous electric field, forcing the upper state (antisymmetric) molecules into the beam axis and the lower state (symmetric) molecules away from the beam axis. With the use of a resonance cavity, the maser becomes self-sustaining and emits radiation at 23.8 GHz. This project evolved further to become the invention of the laser, awarding Townes and his colleagues the Nobel Prize in 1964.

Returning to our dielectric configuration problem, the following chapter is a study of the results from solving Schrodinger's Equation with this double infinite well configuration.

\subsection{Schrodinger Solutions}\label{S:Schro}
Recalling from Section \ref{sec:potentialDerivations}, we can use either Eq \ref{eq:smytheFinal} or Eq \ref{eq:slabImages} as the potential, since we showed the two were equivalent. We again note the inclusion of the multiplicative factor of $1/2$ as explained earlier.

At the midpoint between the two plates, the potential reaches a maximum value, which has relative importance; for energy levels $E$ less than this $U_{max}$, we find bound state energies in pairs, and for $E$ greater than $U_{max}$ we find box states instead, since the particle does not feel the draw of the double well. 

The following figure presents an example of a ground state wavefunction pair (unnormalized) with D = 1.6nm as the gap distance.
\\
\noindent\makebox[\textwidth]{%

\includegraphics[scale=.5]{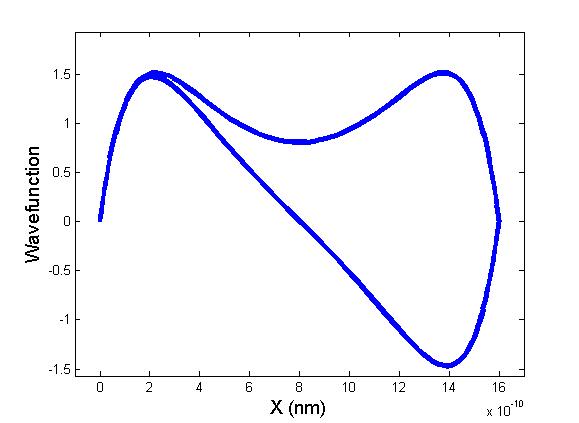}
\label{F:WavefunctionPair}
}

Above, we see the even/bonding lower energy state plotted with the odd/antibonding higher energy state.

In line with the goal of determining plate pressure, we found the first two box and bound energies versus plate gap distance, then subsequently took the spatial derivative to determine the force. We note that the force is repulsive; the binding energies are higher in general for small plate gaps, and thus the tendency is for the plates to separate and lower the energy.

\noindent\makebox[\textwidth]{%

\includegraphics[scale=.4]{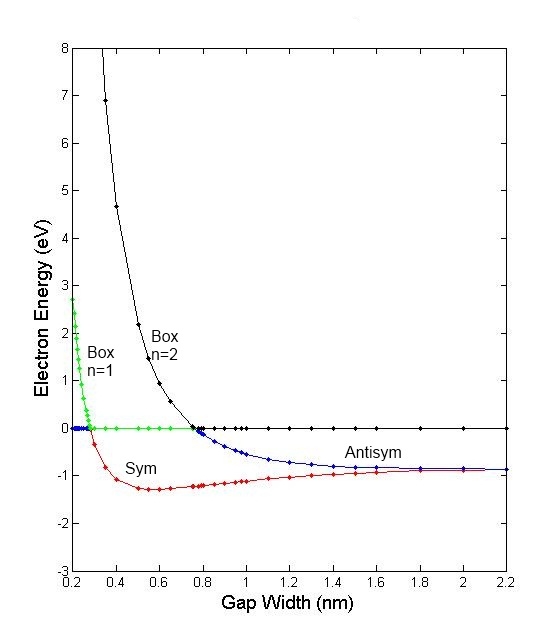}
\label{F:BindingEnergy}

\includegraphics[scale=.4]{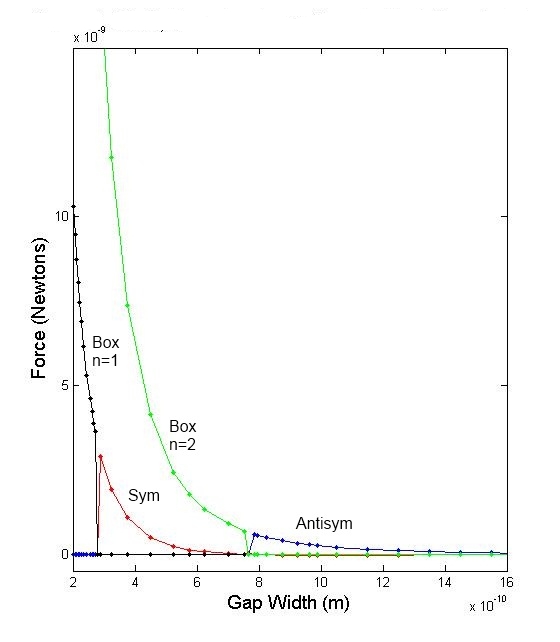}
\label{F:BindingForce}
}

A cardinal result that we can interpret from this graph is the apparent continuity of the box state energies to unique corresponding bound state energies. Indeed, it seems that as the plate gap distance changes, the shapes of the wavefunctions seamlessly transition from even and odd box states to symmetric and antisymmetric bound states, respectively.

\subsection{Comparison of Forces}
However, the repulsive force from the electronic energy is not the only factor in plate repulsion; we recall that we brought charge into the grand picture not only for binding energy purposes, but to create same-sign surface charge on the plates to provide additional repulsion.
Recalling the distance-charge coordinate pairs given by Eq \ref{eq:positions1} and Eq \ref{eq:positions2}, we can determine the pairwise interactions of image charges across the plates to determine the repulsive potential energy, equivalent to the repulsion due to the surface charge on the plates. The interactions with the real point charge in $K_2$ are not counted. The image charges for each plate change their mutual distances as the real point charge travels from one plate interface to the other, resulting in a local maximum of the potential when the point charge is located at the center between the two plates. In the figure below, we show an example of this phenomenon for a gap width of $7.5$ angstroms.

\noindent\makebox[\textwidth]{%
\includegraphics[scale=.4]{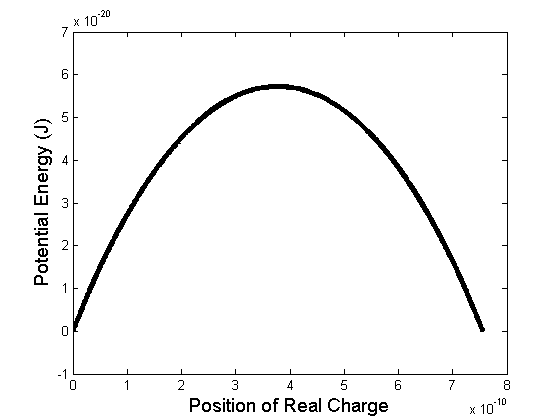}
\label{F:PlatePlateVsChargeLocation}
}

Classically, we would say that the particle would either spend its time evenly near each plate, or that two half charges would be located simultaneously near both plates. However by virtue of quantum mechanics, we cannot say that the particle would be located specifically in either of these configurations. The definition of the squared normalized wavefunction is the probability of locating the particle at each position between the plates. Thus, we quantum mechanically average the plate-plate potential by multiplying by the squared normalized wavefunctions and integrating across the distance between the plates, giving us a scalar for a particular gap width. We then plot this averaged plate-plate interaction energy versus gap width in the figures below, as well as its associated force. Note that the small amount of noise on the force plots is due to numerical differentiation and does not reflect an artifact of the physical system.

\noindent\makebox[\textwidth]{%
\includegraphics[scale=.4]{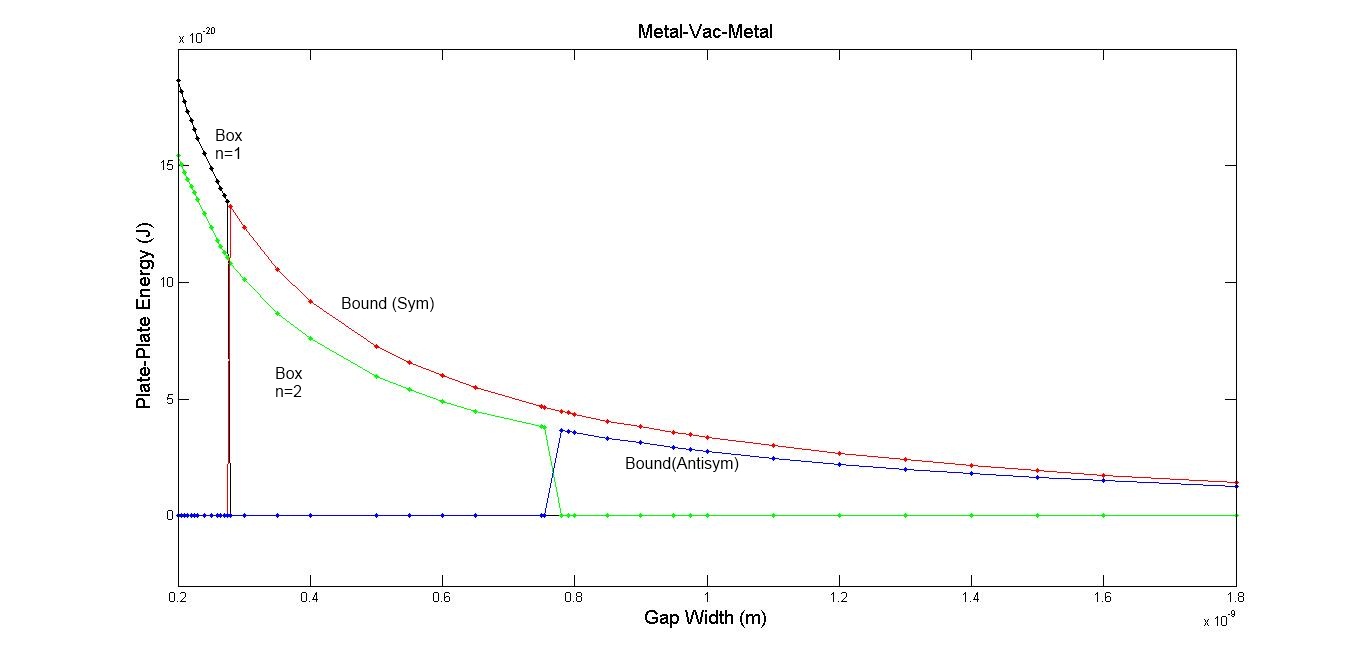}
\label{F:PlateEnergy}
}

\noindent\makebox[\textwidth]{%
\includegraphics[scale=.4]{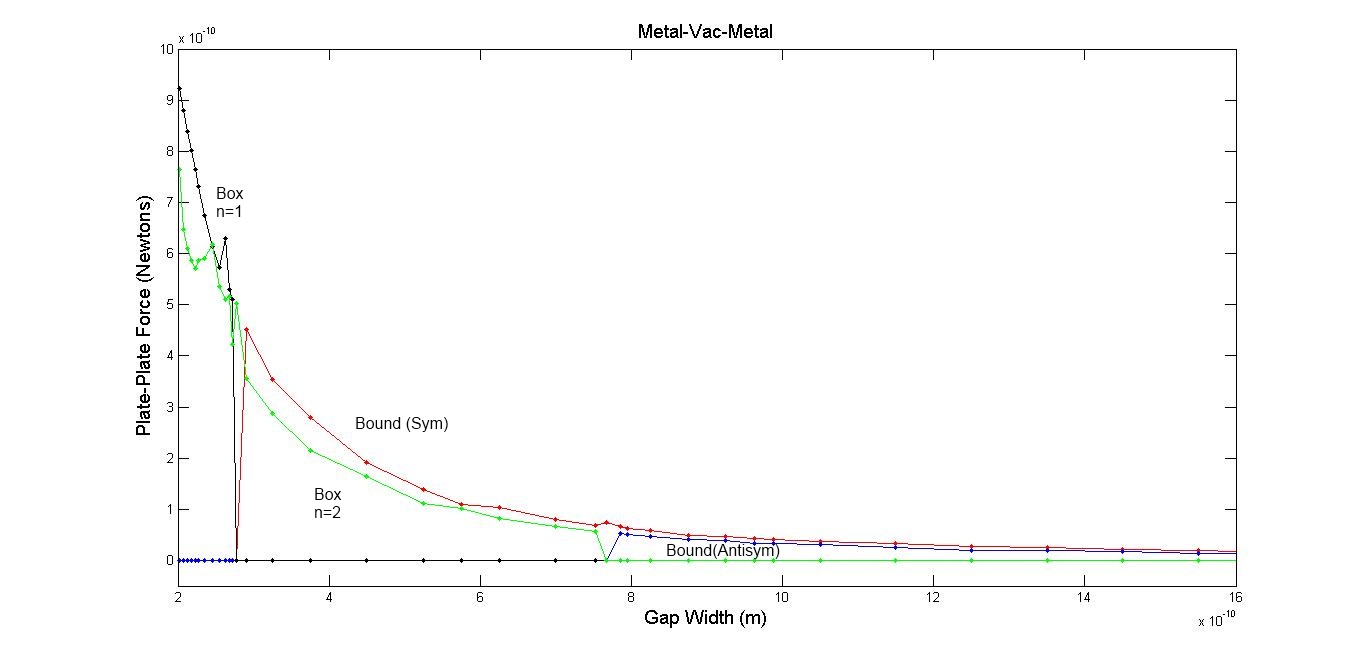}
\label{F:PlateForce}
}

We note that the plate-plate curve associated with the bound symmetric and 1st box state wavefunctions is slightly higher in magnitude than the curve associated with the bound antisymmetric and 2nd box state. This is a logical finding, because the antisymmetric bound state and odd parity box state wavefunctions cross the x-axis at the center; thus the particle is forbidden at the center, which is where the plate-plate image interaction energy is at a maxima, leading to a smaller contribution.

Finally, we add the above electronic energy and the plate-plate interaction energy together to provide us with the total repulsive energy and force on the plates.

\noindent\makebox[\textwidth]{%

\includegraphics[scale=.4]{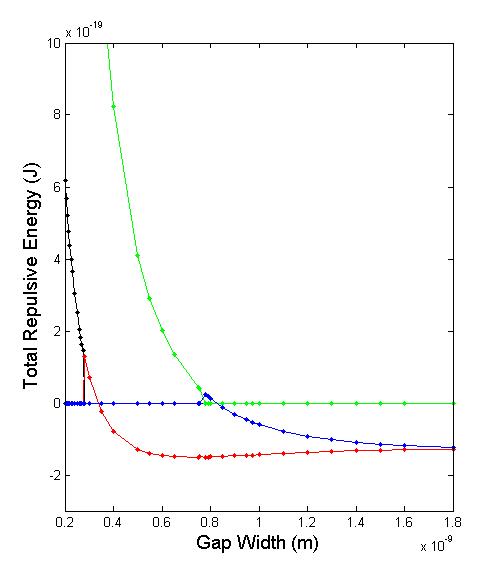}
\label{F:TotalRepulsiveEnergy}

\includegraphics[scale=.4]{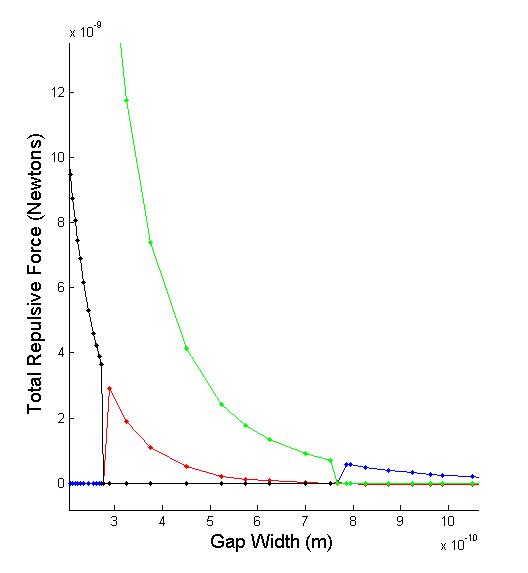}
\label{F:TotalRepulsiveForce}
}

Ideally, we would have only these repulsive forces present, but we find that at close distances, other quantum mechanical repulsive forces are at play, due to the proximity of the plate atoms. The Casimir force, though it decays as a quartic function, is a considerable attractive force in the nanometer distance range. For two parallel plates, the equation is as follows \cite{CasimirOriginal}: 

\begin{equation}
F_{Casimir} = - \frac{\hbar c \pi^2 A}{240 D^4}
\end{equation}

where $A$ is the area of the plate.

Furthermore, the Van Der Waals force between the plates, with a cubic falloff rate, is also a presence at this small scale \cite{CasimirVDW}.

\begin{equation}
F_{VDW} = - \frac{H}{6 \pi D^3}
\end{equation}

where $H$ is the Hamaker constant, which is dependent on the material of the plates. 

For exactly one electron, these two attractive forces dwarf the electronic repulsive force. However, as we increase the number of electrons ($N$) present in the system, the repulsive forces scale independently of the attractive forces. We thus increase N until we approach comparable magnitudes between the the attractive and repulsive forces. As N increases, we see that the plate-plate interaction grows quadratically and the binding energies grow linearly. We then write the total force as a function of N as follows:

\begin{equation}
F_{Total} = N^2 \cdot F_{PlatePlate} + N \cdot F_{Binding} + F_{Casimir} + F_{VDW}
\end{equation}

Where each $F$ term is for that of one electron.

We assume a reasonable experimental area of the plate at $10^{-6}$ to try to determine the number of electrons necessary to counteract the large attractive forces. We find that $3*10^17$ electrons bring the forces to similar magnitude. The following graph is the total energy and force, taking into consideration all repulsive and attractive forces discussed, for $N=3*10^7$ electrons. We are, however, increasing the number of electrons without taking the Pauli Exclusion Principle into account; two electrons cannot have the same quantum numbers, and we must place additional electrons in higher energy states. This is a priority for our future work on this project.

\begin{figure}
\includegraphics[scale=.4]{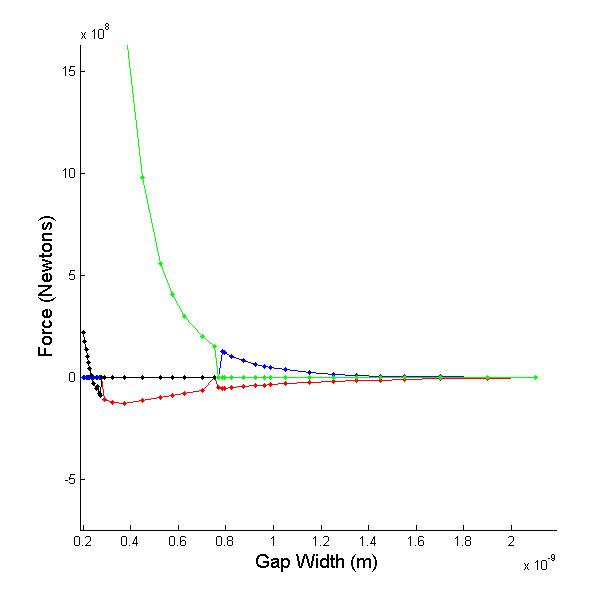}
\label{F:TotalForce}
\end{figure}

With this same configuration and parameters, we can equate the total plate force to the gravitational force, as done in Section \ref{S:particlePlates}. In the below graph, the y-coordinates represent the maximum amount of plate mass able to be levitated at the corresponding gap width. In this sense, any mass-width pair represents an equilibrium point: for a given gap width, a mass too heavy will compress the plate width, which in turn increases the force and tolerates a heavier mass. Similarly, a mass too light for the given gap width will be repelled until the mass is enough to counteract the repulsion. For a fixed mass, it is then possible to construct an oscillator by perturbing the mass from its equilibrium gap width.
In this figure, we consider just the first bound and box state. 

\noindent\makebox[\textwidth]{%
\includegraphics[scale=.4]{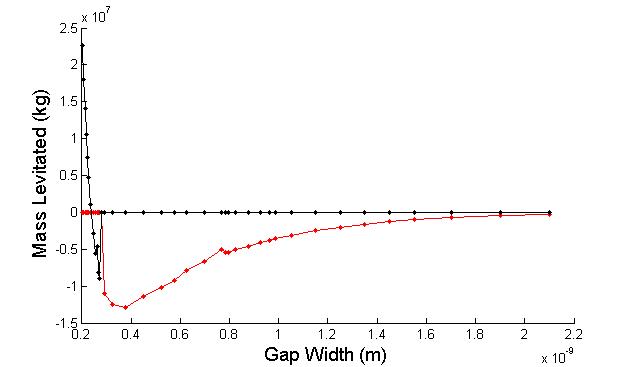}
\label{F:MassLevitated}
}

\section*{Concluding Remarks}
In the course of this research, we were able to verify that the field method of finding the electron potential in Smythe's book \cite{Smythe} is equivalent to the sum of the infinite series from the image charge method. We then utilized these potential functions to explore tunability of the electronic states for different configurations of dielectrics. The first was the use of a vacuum gap to mitigate the band bending phenomenon across a Schottky diode; the results were that the binding energy and Bohr radii are finely tunable by controlling the vacuum gap width. We then showed possible tunability with similar configuration using liquid/solid noble gases in discrete layers between a vacuum and metal half-planes. Finally, we have numerically determined the energy spectrum and wavefunctions of a charge in between two dielectric media, including electrostatic images interactions. We found that the symmetric and antisymmetric bound states are shown to smoothly connect to box states as the gap separation gets smaller. Our next steps are to consider the multi-electron problem and the consequences of the Pauli Exclusion Principle.

\section*{Acknowledgments}
The author is thankful to Prof. Lawandy for his role as mentor and research advisor. The author would also like to thank the Brown University engineering department for consideration of this article in the senior thesis presentations. This research was conducted during the 2011-2012 academic year at Brown University.



\begin{thebibliography}{}

\end{thebibliography}


\begin{thebibliography}{99}
\bibliographystyle{plain}

\bibitem{MolSpec}{G. M. Barrow; \emph{Introduction to Molecular Spectroscopy}; International Ed edition. (McGraw-Hill Inc.,US, 1962)}

\bibitem{Ammonia}{"Double Well"; $http://www.chemsoft.ch/chemed/linbox6.htm$}

\bibitem{Maser}{J. P. Gordon, H. J. Zeiger, and C. H. Townes; "Molecular Microwave Oscillator and New Hyperfine Structure in the Microwave Spectrum of NH3"; Phys. Rev. 95, 282–284 (1954) }

\bibitem{Wavefunctions}{D. J. Griffiths; "Introduction to Quantum Mechanics", Prentice Hall, Upper Saddle River, New 
Jersey 07558 (1995)}

\bibitem{Smythe}{W.R. Smythe; \emph{Static and Dynamic Electricity}; 3rd ed. (McGraw-Hill, New York, 1968).}

\bibitem{CasimirVDW}{G. L. Klimchitskaya, U. Mohideen, V. M. Mostepanenko; "Casimir and van der Waals force between two plates or a sphere (lens) above a plate made of real metals"; Phys. Rev. A 61, 062107 (2000)}

\bibitem{Interfacial}{A. Tugulea, D. Dascalu; "The image-force effect at a metal-semiconductor contact with an interfacial insulator layer"; J. Appl. Phys. 56, 2823 (1984);}

\bibitem{SmytheCheck}{T Sometani; "Image method for a dielectric plate and a point charge"; Eur. J. Phys. Vol: 21, Issue: 6, 549-554 (2000)}

\bibitem{ImagePlane}{M. W. Cole, M. H. Cohen "Image-Potential-Induced Surface Bands in Insulators"; Phys. Rev. Lett. 23, 1238–1241 (1969)}

\bibitem{HydrogenNumerical}{A. Neethiulagarajan, S. Balasubramanian; "On numerical solutions of the radial Schrodinger equation"; Eur. J. Phys. 10 93 (1989)}

\bibitem{Lifshitz}{L. D. Landau, E. M. Lifshitz, L. P. Pitaevskii; \emph{Electrodynamics of Continuous Media}; Butterworth-Heinemann, Jan 1, 1984}

\bibitem{Schottky}{M. Balkanski, R.F. Wallis. "Semiconductor Physics and Applications". Oxford University Press. ISBN 0198517408. (2000)} 

\bibitem{HeliumEpsilon}{C. J. Grebenkemper, J. P. Hagen; "The Dielectric Constant of Liquid Helium"; Phys. Rev. 80, 89–89 (1950)}

\bibitem{CasimirOriginal}{H. B. G. Casimir; \emph{On the attraction between two perfectly conducting plates.}; Communicated at the meeting of May 29, 1948.}

\bibitem{Helium}{M.J.Lea, P.G.Frayne and Y.Mukharsky, "Could we compute with electrons on helium?"; Fortschritte der Physik, 1109 - 112, 48 (2000)}

\bibitem{InversionAmmonia}{Inversion Spectrum of Ammonia, University of Washington. $http://courses.washington.edu/phys432/NH3/ammonia_inversion.pdf$}

\bibitem{HeliumThickness}{K. R. Atkins; "Liquid Helium Films. I. The Thickness of the Film"; Proceedings of the Royal Society of London. Series A, Mathematical and Physical Sciences; Vol. 203, No. 1072, (1950)}

\bibitem{ArgonThickness}{J. Chomaa, J. Górkab, M. Jaroniec; "Mesoporous carbons synthesized by soft-templating method: Determination of pore size distribution from argon and nitrogen adsorption isotherms"; Volume 112, Issues 1–3, 573-579 (2008)}

\bibitem{GaAs}{R.E. Neidert, "Dielectric constant of semi-insulating gallium arsenide"; Electronics Letters, Volume 16,  Issue 7 (1980)}

\bibitem{InSb}{J.R. Dixon Jr., J.K. Furdyna; "Measurement of the static dielectric constant of the InSb lattice via gyrotropic sphere resonances"; Solid State Communications, Volume 35, Issue 2,  195-198 (1980)}

\bibitem{GaAsEffectiveMass}{L.C. Barcus, A. Perlmutter, J. Callaway; "Effective Mass of Electrons in Gallium Arsenide"; Physical Review, Volume 111, Issue 1, pp. 167-168 (1958)}

\bibitem{InSbEffectiveMass}{M. Levinshtein, S. Rumyantsev, M. Shur, ed.; "Handbook Series on Semiconductor Parameters"; World Scientific, London, 1996, Volume 1, pp. 191-213 (1996)}

\end{thebibliography}
\end{document}